\documentclass[traditabstract]{aa}
\usepackage{verbatim} 
\usepackage{amsmath}
\usepackage{amssymb}
\usepackage{graphicx}
\usepackage{booktabs}
\usepackage[authoryear]{natbib}
\usepackage[colorlinks=true,linkcolor=black,urlcolor=blue,citecolor=blue]{hyperref}
\usepackage{xspace}
\usepackage{siunitx}

\usepackage[nolist]{acronym}
\usepackage{lipsum}

\makeatletter

\usepackage{txfonts}\usepackage{twoopt}
\bibpunct{(}{)}{;}{a}{}{,}
\usepackage{enumitem}
\usepackage{xspace}

\newcommand{\msol}{M_{\odot}}
\newcommand{\rsol}{R_{\odot}}

\newcommand{\eg}{e.g.\@\xspace}

\newcommand{\ie}{i.e.\@\xspace}

\newcommand{\arepo}{\textsc{arepo}\xspace}
\newcommand{\mesa}{\textsc{MESA}\xspace}
\titlerunning{Formation of sdB-stars via common envelope ejection by substellar companions}
\authorrunning{M. Kramer et al.}
\makeatother
\begin{document}
\title{Formation of sdB-stars via common envelope ejection by substellar companions}
\author{M.~Kramer\inst{\ref{HITS}}\thanks{\email{manuel.kramer@h-its.org}} \and
  F.~R.~N.~Schneider\inst{\ref{HITS},\ref{ARI}} \and
  S.~T.~Ohlmann\inst{\ref{RZG}} \and
  S.~Geier\inst{\ref{Potsdam}} \and
  V.~Schaffenroth\inst{\ref{Potsdam}} \and
  R.~Pakmor\inst{\ref{MPA}} \and
  F.~K.~R{\"o}pke \inst{\ref{HITS},\ref{ITA}}}
\institute{%
    Heidelberger Institut f\"{u}r Theoretische Studien,
    Schloss-Wolfsbrunnenweg 35, 
    69118 Heidelberg, Germany\label{HITS}
  \and
    Zentrum f\"ur Astronomie der Universit\"at Heidelberg,
    Astronomisches Rechen-Institut, 
    M\"{o}nchhofstr. 12-14, 
    69120 Heidelberg, Germany\label{ARI}
  \and
    Max Planck Computing and Data Facility,
    Gießenbachstr. 2,
    85748 Garching, Germany\label{RZG}
  \and
    Institut f{\"u}r Physik und Astronomie,
    Universit{\"a}t Potsdam, Haus 28,
    Karl-Liebknecht-Str. 24/25,
    14476 Potsdam-Golm, Germany\label{Potsdam}
  \and
    Max-Planck-Institut für Astrophysik,
    Karl-Schwarzschild-Str. 1,
    85748 Garching, Germany\label{MPA}
  \and
    Zentrum f\"ur Astronomie der Universit\"at Heidelberg,
    Institut f\"ur Theoretische Astrophysik, 
    Philosophenweg 12,
    69120 Heidelberg, Germany\label{ITA}
}

\date{Received xxx / Accepted yyy}
\abstract{Abstract}
\abstract{Common envelope (CE) phases in binary systems where the
  primary star reaches the tip of the red giant branch are discussed
  as a formation scenario for hot subluminous B-type (sdB) stars. For
  some of these objects, observations point to very low-mass
  companions. In hydrodynamical CE simulations with the moving-mesh
  code \arepo, we test whether low-mass objects can successfully
  unbind the envelope. The success of envelope removal in our
  simulations critically depends on whether or not the ionization
  energy released by recombination processes in the expanding material
  is taken into account. If this energy is thermalized locally,
  envelope ejection eventually leading to the formation of an sdB star
  is possible with companion masses down to the brown dwarf range. For
  even lower companion masses approaching the regime of giant planets,
  however, envelope removal becomes increasingly difficult or
  impossible to achieve. Our results are consistent with current
  observational constraints on companion masses of sdB stars. Based on
  a semianalytic model, we suggest a new criterion for the lowest
  companion mass that is capable of triggering a dynamical response of
  the primary star thus potentially facilitating the ejection of a
  common envelope. This gives an estimate consistent with the findings
  of our hydrodynamical simulations.}

\maketitle
%
%
%
%
\section{Introduction}
\label{sec:introduction}

\acp{sdB} stars are helium-core-burning stars that contain almost no
hydrogen.  They reach hot surface temperatures of about
\SIrange{2e4}{4e4}{K}, which places them on the blue end of the
horizontal branch \citep{heber1986a}. To form \ac{sdB} stars, almost
all of the hydrogen envelope of the progenitor must be removed at the
same time as the helium ignition is triggered in the core. When this
process commences, it thus has most likely evolved to the tip of the
\ac{RGB}.

A natural mechanism for removing the hydrogen envelope is the
interaction with a binary companion.  About 50\% of
sun-like stars evolve alongside a companion and this fraction is even
higher for more massive stars \citep{duchene2013a, moe2017a}. When one star in a
close binary system reaches the \ac{RGB}, it expands rapidly, thus
overfilling its \ac{RL}, and can trigger unstable mass transfer. If
the receiving companion cannot accrete all of this material, it will
be engulfed and a \ac{CE} is formed \citep{paczynski1976a} around the
two compact stellar cores. These cores spiral inwards and transfer
angular momentum and energy to the envelope material. As a result, the
envelope expands and might be partially or even completely ejected
from the system \citep{ivanova2013a}. The separation between the cores
is greatly reduced and a close binary forms.

Observations have indeed shown that 40\% to 70\% of the
single-lined \ac{sdB} stars exist in close binary systems with periods
ranging from \SIrange{0.03}{10}{d} \citep{maxted2001a}. This strongly
suggests a previous \ac{CE} phase, in which the orbital separation is
reduced and the \ac{RG} progenitor loses most of his hydrogen-rich
envelope material in the interaction with its companion. Surprisingly,
in recent surveys, several \ac{sdB} binaries with companions in the
\ac{BD} regime have been found \citep{geier2011a, schaffenroth2014a,
  schaffenroth2015a}. This raises the question of whether a \ac{CE}
interaction with such low-mass companions can indeed trigger
successful envelope ejection. Based on simple estimates,
\citet{soker1998a} and \cite{nelemans1998a} argue that companions with
masses lower than about $10^{-2}\, \msol$ evaporate or lose their
mass in \ac{RL} overflow before completely ejecting the envelope
material.

Such estimates bear large uncertainties and call for a closer
investigation. The dynamics of \ac{CEE} can only be captured
self-consistently in three-dimensional hydrodynamical
simulations. With the wide range of spatial scales involved and the
need to follow the system over many orbits, these pose substantial
challenges to numerical approaches. Smoothed-particle hydrodynamics
(SPH) offers a way to account for the ``Lagrangian nature'' of the
problem, but it usually lacks spatial resolution in the dilute stellar
envelopes. An improvement are moving mesh techniques \citep[][Sand et
  al., in prep.]{ohlmann2016a, ohlmann2016b, prust2019a}, that combine
the efficiency of (nearly) Lagrangian methods with the accuracy of
grid-based hydrodynamics solvers. Despite recent progress in numerical
techniques and available computational resources, a fundamental
question of \ac{CEE} remains unanswered: How is the envelope ejected?
If driven by the release of orbital energy only, the ejection remains
incomplete in all published \ac{CE} simulations. Additional physics
seems to be required for a successful envelope removal. The ionization
energy stored in the envelope will be released by recombination
processes provided that the material expands sufficiently. If
thermalized locally, this energy leads to further unbinding of
material \citep{nandez2015a, prust2019a} and a complete envelope
ejection seems possible.

So far, no hydrodynamic \ac{CE} simulations have been carried out in
the context of the formation of \ac{sdB} stars.  Based on such
simulations, the work presented here aims to determine if substellar
companions are sufficient to trigger a significant unbinding of the
envelope material in cases where the primary star is at the tip of its
\ac{RGB}.  In the subsequent sections, we explain the methods and the
setup of our simulations (Sect.~\ref{sec:methods}) and present their
results (Sect.~\ref{sec:results}). Based on these, we develop a
semi-analytic model for the inspiral of the stellar cores
(Sect.~\ref{sec:minimum-mass}), discuss the fate of the companion and
compare our results to observations (Sect.~\ref{sec:discussion})
before concluding (Sect.~\ref{sec:conclusion}).

%
%
\section{Methods}
\label{sec:methods}

Following the work of \cite{ohlmann2016a, ohlmann2016b, ohlmann2017a},
we employ the moving-mesh magnetohydrodynamics code \arepo
\citep{springel2010a, pakmor2011d, pakmor2013b} to simulate the
\ac{CE} phase in a system composed of a primary star at the tip of its
\ac{RGB} and a compact low-mass companion. This code is particularly
well-suited for this task because of its shock capturing abilities and
the excellent conservation of angular momentum and energy. It allows
for arbitrary refinement criteria to achieve higher resolution in
specific areas. For most of our simulations, we use the OPAL \ac{EoS}
\citep{rogers1996a, rogers2002a}. It accounts for ionization effects
and allows us to track the release of recombination energy. This
energy is in our current implementation assumed to be thermalized
locally. For comparison, one simulation is carried out with an ideal
gas \ac{EoS}.

\subsection{RG star model}

With the stellar evolution code \mesa \citep{paxton2013a,paxton2015a}
in version 6208, we create a suitable \ac{RG} progenitor model as the
primary star (denoted with subscript 1 in the following) for the
subsequent binary simulations. A \(1 \, \msol\) zero-age main sequence
star is evolved until the tip of the red giant branch, applying
default \mesa settings. The metallicity is set to $Z = 0.02$ and the
Reimers prescription with $\eta = 0.5$ is used for \ac{RG} winds. Due
to mass loss via winds, we reach a stellar model that has a total mass
of $ M_1 = M_\mathrm{RG} = 0.77 \, \msol $, a radius of $R_1 =
R_\mathrm{RG} = 173 \, \rsol$, a core mass of $ M_\mathrm{core} =
0.47\,\msol $ and a envelope mass of $ M_\mathrm{env} = 0.30\,\msol $.

The \ac{1D} \mesa profile is then mapped onto a \ac{3D} grid following
the procedure described in \citet{ohlmann2017a}. We cut out the core
at five percent of the radius of the \mesa model and replace it with a
point mass that only interacts gravitationally, henceforth called
``core particle''. To obtain a stellar model in \ac{HSE}, where
\begin{align}\label{eq:hse}
  \rho \vec{g} = \nabla P,
\end{align}
a modified Lane-Emden equation is solved to create a profile with a
smooth transition between the core and the envelope
\citep{ohlmann2017a}.

\subsection{Relaxation} \label{sec:relaxation}

\begin{figure}
  \resizebox{\hsize}{!}{\includegraphics{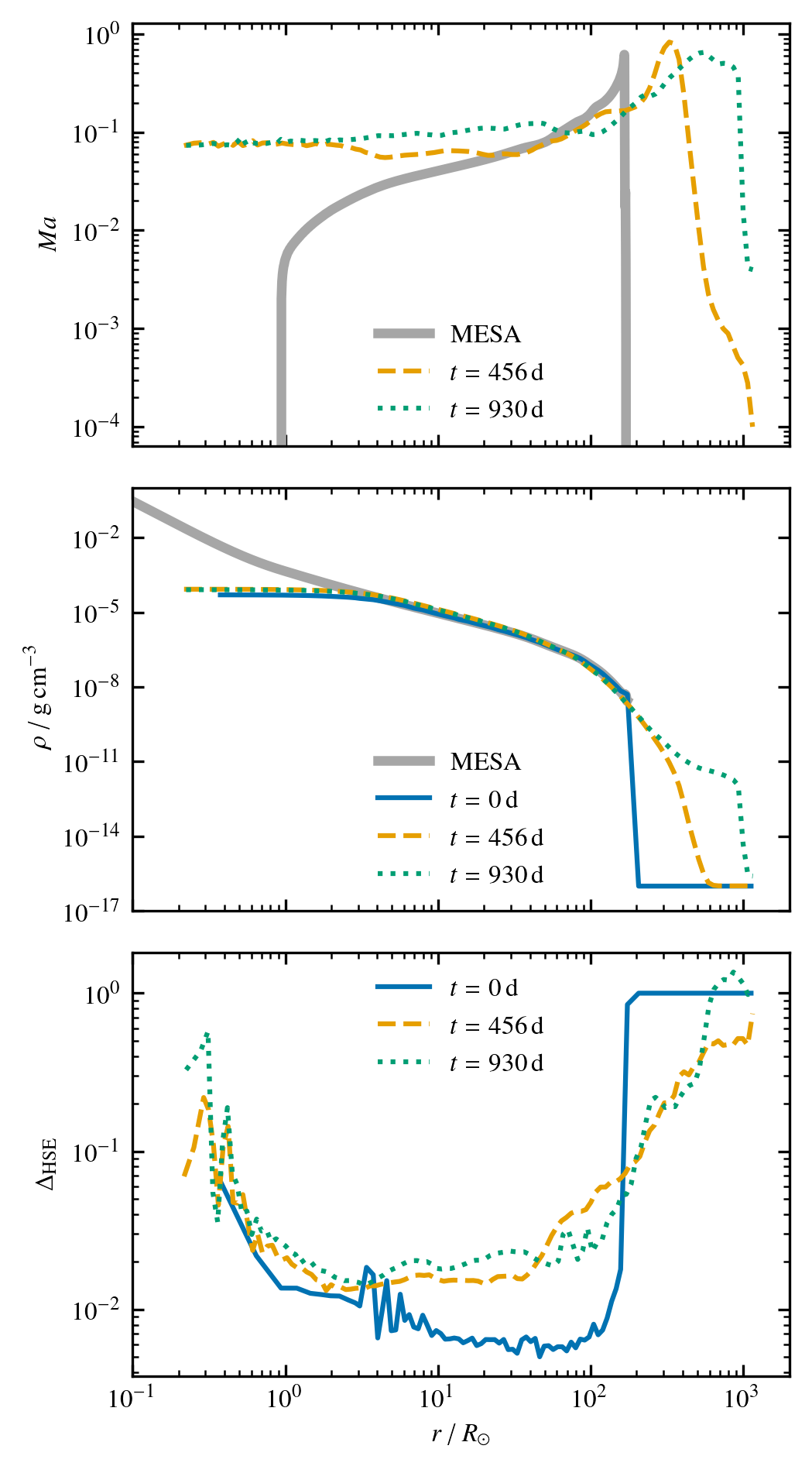}}
  \caption{Mach number \(Ma\), density \(\rho\) and relative deviation
    from \ac{HSE} \(\Delta_\mathrm{HSE}\) according to
    Eq.~(\ref{eq:delta_hse}) over radius of the relaxed giant at the
    tip of the \ac{RGB}. The quantities are binned over the radius and
    averaged in shells.}
  \label{fig:relaxation}
\end{figure}

Because the spatial resolution is coarser in our \ac{3D} setup than in
the \ac{1D} stellar evolution model and because the employed \ac{EoS}
is not identical, spurious motions in the mapped stellar structure
arise. To restore \ac{HSE}, we carry out a relaxation run of the
mapped stellar model for 10 dynamical timescales $t_\mathrm{dyn}$,
which corresponds to \SI{930}{d}.  Velocities are damped by a constant
factor during the first 2 \(t_\mathrm{dyn}\) and then constantly
reduced until we reach \(t = 5 \, t_\mathrm{dyn}\). For the remaining
5 \(t_\mathrm{dyn}\), the damping is completely shut off which allows
us to check if the star stays stable in the relaxed configuration.

In the top panel of Fig.~\ref{fig:relaxation}, the Mach number, the
ratio between the absolute value of the local fluid velocity $v$ and
the speed of sound $c_\mathrm{s}$, $\mathit{Ma} \equiv
v/c_\mathrm{s}$, in the mapped stellar model is plotted over the
radius at different times.  The inner part of the star stays subsonic
with $\mathit{Ma} \sim 0.1$.  This shows that spurious velocities are
damped successfully and the star's envelope settles into a stable
state. For technical reasons, the grid outside the star cannot be
empty but has to be filled with low-density material. We chose a
uniform background density of \(\rho_\mathrm{bg} = 10^{-16} \,
\mathrm{g} \, \mathrm{cm}^{-3}\), and the material is not in
\ac{HSE}. Consequently, in these regions, Mach numbers above $0.5$
occur, but because these flows contain only \(\SI{6.3e-4}{} \msol\),
they are irrelevant for the dynamics of the stellar envelope.

The middle panel of Fig.~\ref{fig:relaxation} shows that the density
profile does not change over time after we stop the damping. We
observe some expansion of surface material. This is caused by the
steep initial pressure gradient, that cannot be fully resolved, which
makes it impossible to fulfill the condition of \ac{HSE}
(\ref{eq:hse}) for the original profile. Therefore, the relaxed
profile settles into a new equilibrium with shallower surface
gradients. Still, no mass is lost from the system and the original
profile of the \mesa output is well represented in the inner parts of
the star, where most of the mass is concentrated.

In the bottom panel of Fig.~\ref{fig:relaxation}, the relative
difference between both sides of the \ac{HSE} equation (\ref{eq:hse}),
\begin{align}\label{eq:delta_hse}
  \Delta_\mathrm{HSE} \equiv \frac{|\rho \vec{g} - \nabla P|}{\max
  \left( |\rho \vec{g}|, |\nabla P| \right)},
\end{align}
is shown. Throughout most of the envelope, deviations from \ac{HSE}
stay at low values of $\Delta_\mathrm{HSE} \approx 0.02$. Near the
center, close to the core particle, the error increases due to the
slight decrease in density, as well as close to the surface due to the
expansion.

We compute a sphere centered on the core particle that contains
\SI{99.9}{\percent} of the mass of the initial \ac{RG} to define a
final radius of \(R_1 = R_\mathrm{RG} = 118 \, \rsol \) at the end of
the relaxation.

This shows that the relaxed model represents a star at the tip of the
\ac{RGB}. It stays stable over sufficiently many dynamical timescales
to simulate the subsequent \ac{CE} phase in a binary system.

\subsection{Binary simulations}

To setup our binary simulations, we place a compact companion --
denoted with subscript 2 in the following and technically realized as
a second core particle of mass \(M_2\) -- at a orbital separation
\(a\) that corresponds to 80\% of the \ac{RL} radius. The binary
components are placed on a Kepler orbit at a frequency of
\begin{align}
    \Omega = \sqrt{\frac{G (M_1 + M_2)}{a^3}}, 
\end{align}
where $G$ is the gravitational constant. To facilitate the inspiral,
we impose a corotation factor of \(\chi = 0.95\), so that initial
velocity of the envelope is given by:
\begin{align}
   \vec{v}_\mathrm{env} = \chi \left( \vec{\Omega} \times [\vec{r} -
     \vec{r}_\mathrm{core}] \right)
\end{align} 
Here, \(\vec{r}_\mathrm{core}\) is the position of the \ac{RG} core
and \(\vec{\Omega}\) points into the \(z\)-direction.

In our simulations, we solve for full magnetohydrodynamics. Following
\cite{ohlmann2016b}, the magnetic field of the \ac{RG} star is
initialized as dipole along the $z$-axis with $10^{-6}\,\mathrm{G}$ at
the pole. In our current treatise, however, we focus on the
hydrodynamical evolution. The magnetic fields in our simulations are
dynamically irrelevant and will be discussed in a forthcoming
publication.

\section{Results} \label{sec:results}

In this section, we present the results of our \ac{3D} hydrodynamical
\ac{CE} simulations. All are based on the same initial \mesa model for
the \ac{RG} primary star. Some general features of the dynamics are
described in Sect.~\ref{sec:reference_run} based on a ``reference
simulation'' with a companion mass of \(M_2 = 0.08 \, \msol\). We then
vary \emph{model parameters} of the reference run independently: the
spatial resolution around the core particles
(Sect.~\ref{sec:resolution}) to test numerical convergence of the
simulation and the \ac{EoS} (Sect.~\ref{sec:eos}) to investigate the
effect of recombination energy release on envelope ejection. Based on
the results of these runs, we carry out simulations, which explore the
effect of the most important \emph{physical parameter} of the systems
under consideration -- the mass of the companion -- in a setup
otherwise identical to the reference run
(Sect.~\ref{sec:companion}). Finally, we present a semianalytic model
that yields a new criterion for determining the lowest companion mass
that is still capable of triggering envelope ejection
(Sect.~\ref{sec:minimum-mass}).

\subsection{Reference simulation} \label{sec:reference_run}

\begin{figure}
  \resizebox{\hsize}{!}{\includegraphics{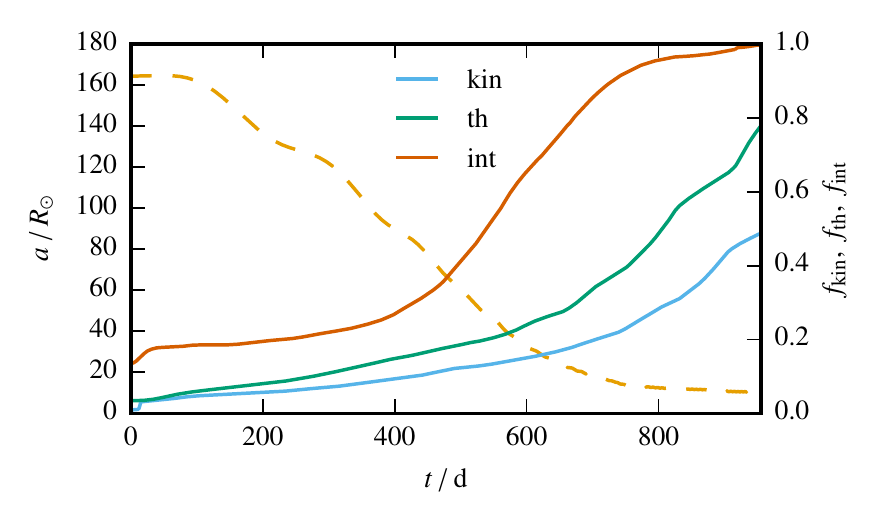}}
  \caption{Orbital separation $a$ between companion and core of the
    \ac{RG} over time $t$ for a companion mass of $0.08 \, \msol$
    (dashed, left axis) and ejected mass fractions $f$ according to
    the three criteria based on the kinetic (``kin''), thermal (``th),
    and internal (``int'') energy as defined in the text (solid, right
    axis).}
  \label{fig:dist_unb}
\end{figure}

For the reference run we choose a companion mass of $M_2 = 0.08 \,
\msol$, implying a mass ratio of the companion to the primary star of
$q \equiv M_2/M_1 = 0.01$. The companion is placed at a distance of
$a_\mathrm{i} = 164 \, \rsol$ to the center of the \ac{RG} (the
initial period at the start of our simulation is $P_\mathrm{i} =
\SI{329}{\day}$) and the OPAL \ac{EoS} is applied. The companion
spirals in, thereby ejecting a large fraction of the envelope. We
follow this process in our simulation for about \SI{1000}{d}. During
the complete run, the relative error amounts to 0.6\% in the total
energy and to 1.3\% in angular momentum.

In Fig.~\ref{fig:dist_unb}, the orbital separation \( a \) and
fraction of unbound mass over time are shown. At $t = \SI{161}{d}$,
the rate of orbital decay \(\dot{a}/a\) surpasses 0.1 and thus
initializes the phase of rapid inspiral, that stops at $t =
\SI{695}{d} $ at a orbital separation of $ a \approx 20 \, \rsol
$. The separation decreases at a slower rate until it reaches $ a =
10.4 \, \rsol$ at the end of the simulation.

\begin{figure*}[t]
  \resizebox{\hsize}{!}{\includegraphics{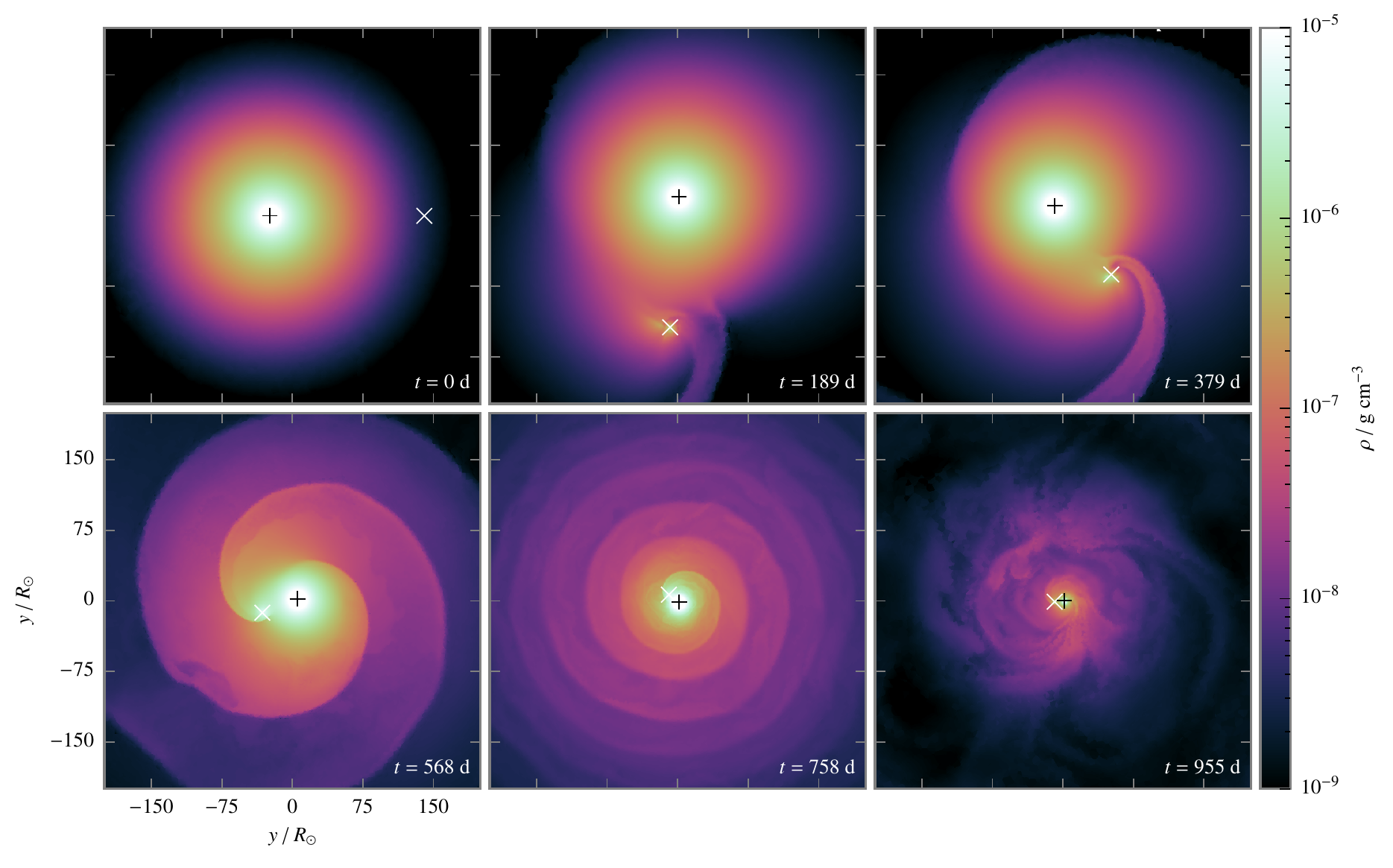}}
  \caption{{Time series of density snapshots in the orbital plane at
      different times.}  The positions of the cores of the \ac{RG}
    primary and the $0.08 \, \msol$ companion are marked by an
    \(\times\) and + respectively. Each frame is centered on the
    center of mass of the binary system.}
  \label{fig:time_series_rho}
\end{figure*}

In Fig.~\ref{fig:time_series_rho}, density slices through the orbital
plane at different times are shown. During the first orbit, the
structure of the \ac{RG} remains almost unperturbed. Between $ t =
\SI{0}{d} $ and $ t \sim \SI{189}{d} $, the companion accumulates mass
in its wake and forms a tidal arm in the envelope material. The
inspiral enters a faster phase and shear instabilities emerge at the
edge of the tidal arm. From $ t \sim \SI{600}{d} $ to $ t \sim
\SI{750}{d} $, a layered structure emerges with shear instabilities
between the adjacent layers. At the end of the simulation at $ t =
\SI{955}{d} $, the initial structure of the \ac{RG} is completely
disrupted and a large fraction of the initial envelope has been
removed.

In Fig.~\ref{fig:dist_unb}, the fraction of unbound mass over time is
depicted.  We apply three different criteria to determine if mass is
unbound, all based on the energy budget of the envelope material: The
``kinetic energy criterion'' counts mass as unbound if
\(e_{\mathrm{kin}}+e_{\mathrm{pot}}>0\), the ``thermal energy
criterion'' if
\(e_{\mathrm{kin}}+e_{\mathrm{pot}}+e_{\mathrm{th}}>0\), and the
``internal energy criterion'' if
\(e_{\mathrm{kin}}+e_{\mathrm{pot}}+e_{\mathrm{int}}>0\), where
\(e_{\mathrm{kin}}\), \(e_{\mathrm{pot}}\), \(e_{\mathrm{th}}\), and
\(e_{\mathrm{int}}\) denote the kinetic, the potential gravitational,
the thermal, and the internal energy of the gas, respectively. The
internal energy includes both the thermal and the ionization energy of
the gas. These criteria provide different estimates of the unbound
mass. While the potential energy criterion regards material as unbound
only if its kinetic energy has overcome the gravitational binding
energy and thus provides a conservative estimate, the other energy
components may ultimately be converted into kinetic energy and thus
contribute to mass ejection, although this has not happened yet at the
instant of measurement. We will refer to the ratio of unbound mass to
the initial envelope mass $M_\mathrm{env}$ under the respective
criterion as \(f_{\mathrm{kin}}\), \(f_{\mathrm{th}}\), and
\(f_{\mathrm{int}}\).

The initial jump in the unbound mass for both the thermal and the
internal energy criterion is due to the recalculation of the energy
contribution when placing the core particle representing the companion
onto the grid containing the relaxed progenitor star.  During the
inspiral, the unbound masses determined with the kinetic and the
thermal energy criterion both increase at a low rate, that grows after
the orbit is stabilized. This is expected, because the ionization
energy of the gas is only converted into thermal and kinetic energy
when the envelope has significantly expanded and cooled. Assuming that
all of the ionization energy will be used eventually, we obtain a
fraction of unbound mass of $ f_\mathrm{int} = 99.7\%$.  The thermal
energy criterion yields $ f_\mathrm{th} = 77.8\%$ of unbound gas and
is still increasing steeply the end of the run, because ionization
energy is still being converted into thermal energy. This strongly
suggests that further mass unbinding will take place. Consequently,
even low mass companions appear to suffice to completely unbind the
envelope of the \ac{RG} and form \ac{sdB} systems.

\subsection{Resolution study} \label{sec:resolution}

\begin{table}[htbp]
    \caption{{Overview of the results of the resolution runs.}}
    \label{tab:resolution}
    \centering
    {\def\arraystretch{1.2}
    \begin{tabular}{ccccccc} 
      \hline\hline
      {$n_{\mathrm{c}}$} & {$a_\mathrm{f} / \rsol$} & {$P_{\mathrm{f}} /
      \mathrm{d}$} & {$ \Delta E_\mathrm{rel} $} &
      {$\Delta J_\mathrm{rel}$} & {$f_\mathrm{th}$} & {$N_\mathrm{f} / 10^{6}$}
      \\ 
      \hline 
        40 & 10.4 & 5.3 & 0.6\% & 1.1\% & 77.8\% & 2.95 \\
        30 & 10.7 & 5.4 & 0.5\% & 0.9\% & 78.2\% & 2.65 \\ 
        20 & 53.7 & 50.4 & 2.2\% & 0.8\% & 71.2\% & 2.32 \\
        10 & 110.5 & 174.5 & 2.9\% & 0.2\% & 90.5\% & 2.14 \\ 
	\hline 
  \end{tabular}}
  \tablefoot{Quantities defined in the text.}
\end{table}

\begin{figure}
  \resizebox{\hsize}{!}{\includegraphics{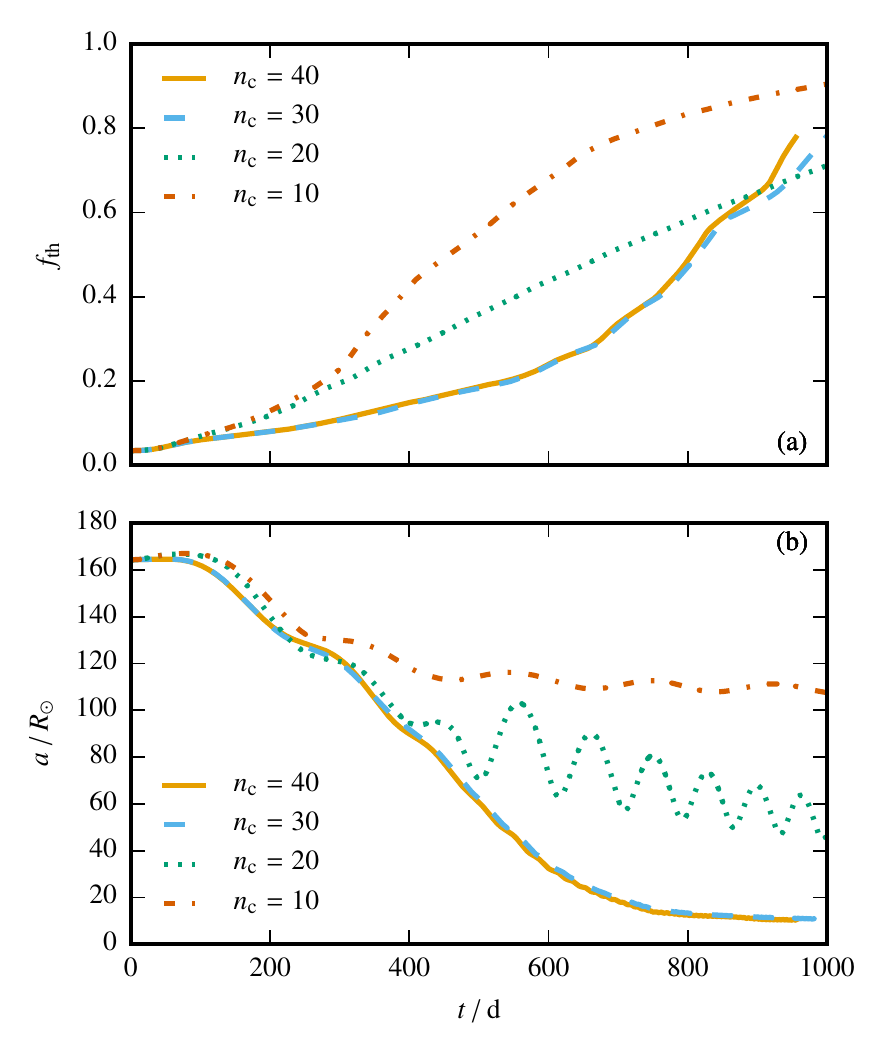}}
  \caption{The upper panel shows the fraction of unbound mass
    $f_\mathrm{th}$ under the thermal energy criterion over time. The
    legend denotes the imposed number of cells $n_\mathrm{c}$ per
    softening length. In the lower panel, the separation between the
    core particles is shown.}
  \label{fig:resolution}
\end{figure}

For the \ac{CE} dynamics, the transfer of orbital energy and angular
momentum from the cores to envelope gas is critical. This proceeds
around the cores. It is therefore necessary to sufficiently resolve
the regions around the core particles in our simulations. As a test,
we conduct a number of simulations with varying spatial
resolution. When combining a grid-based representation of matter with
point particles, the gravitational potential has to be softened
\citep{springel2010a}. This is the case in our simulations and we set
the softening length to \(h = 3.1 \, \rsol\). The initial total number
of cells in our simulation domain is approximately \SI{2.4e6}{}. These
parameters are applied to all simulations presented in this
work. Using a special refinement criterion of \arepo, we vary the
number of cells per softening length \(n_\mathrm{c}\) around the core
particles between 10 and 40 and compare the evolution of the
system. Except for \(n_\mathrm{c}\), we use setups identical to the
reference simulation of Sect.~\ref{sec:reference_run}. We summarize
the convergence test runs in Table~\ref{tab:resolution}. Because the
refinement criterion produces additional grid cells around the core
particles, the final total number of cells $N_\mathrm{f}$ in the
simulation domain is larger by 37\% for the highest-resolved run
compared to that with the lowest resolution (see
Table~\ref{tab:resolution}). This implies a growth in computational
cost and for this reason the run with \(n_\mathrm{c} = 40\) terminates
at \(t = \SI{955}{d}\), while the other three extend to $t =
\SI{1000}{d}$.

In the lower panel of Fig.~\ref{fig:resolution}, the orbital
separation between the core particles is plotted over time. The final
separations $a_\mathrm{f}$ and periods $P_\mathrm{f}$ are given in
Table~\ref{tab:resolution}. The results of the simulations with
$n_\mathrm{c} = 40$ and $n_\mathrm{c} = 30$ and the two simulations
with $n_\mathrm{c} = 20$ and $n_\mathrm{c} = 10$ are qualitatively and
quantitatively different. Overall, the two highest-resolved
simulations are similar in their orbital evolution. The values of the
final unbound mass fractions according to the thermal energy criterion
and also the final orbital parameters are very close. This indicates
that they are numerically converged.  Both 10 and 20 cells per
softening length are insufficient to capture the orbital energy and
angular momentum transfer between the companion and the surrounding
cells, which results in incomplete inspirals. More mass is unbound
earlier in the simulations with low resolution, probably due to
under-resolved gravitational interaction of the core particles. This
can lead to spurious velocities that inject kinetic energy.  It is
interesting to note that the lowest-resolved simulation shows the
largest mass ejection.  This emphasizes the danger of wrong
conclusions drawn from under-resolved \ac{CE} simulations and
underlines the necessity of through convergence studies.

All simulations conserve angular momentum and total energy relatively
well, which is a necessary prerequisite for meaningful physical
results, but, as demonstrated here, not a guarantee for numerical
convergence.  Table~\ref{tab:resolution} summarizes the relative
energy errors $\Delta E_\mathrm{rel} = \frac{1}{E_0}| E_0 -
E_\mathrm{f} |$, with $E_0$ and $E_\mathrm{f}$ denoting the initial
and final values of total energy in our simulation domain, and the
analogously defined relative errors in total angular momentum $\Delta
J_\mathrm{rel} = \frac{1}{J_0} | J_0 - J_\mathrm{f} |$.

From our test simulations, it is clear that the gravitational
softening length has to be resolved with at least 30 cells. To be on
the safe side, we opt for the highest resolution with $n_\mathrm{c} =
40$ in all simulations presented below. This ensures numerically
converged results with favorable angular momentum and energy
conservation.

\subsection{The influence of ionization energy} \label{sec:eos}

\begin{figure}
  \resizebox{\hsize}{!}{\includegraphics{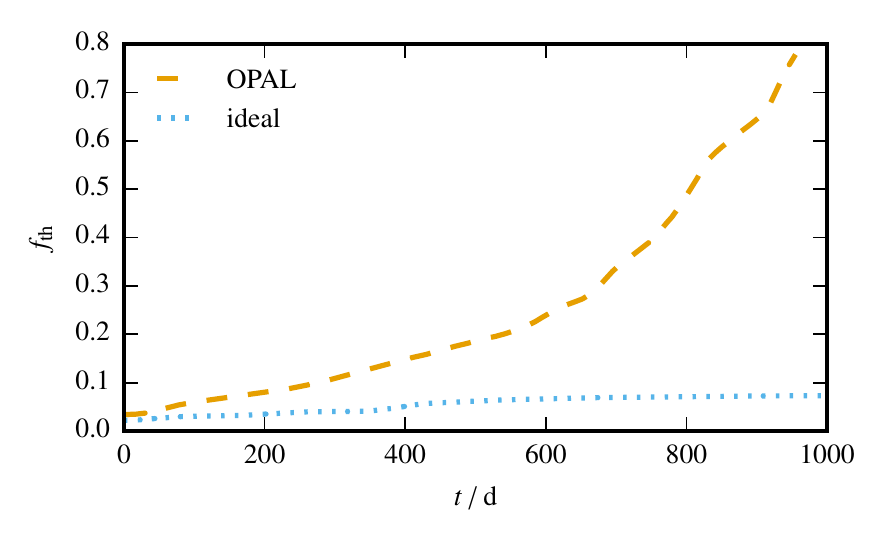}}
  \caption{Evolution of the fraction of the unbound mass
    $f_\mathrm{th}$ assuming the thermal energy criterion for
    simulations with different \ac{EoS}.}
  \label{fig:eos}
\end{figure}

To analyze the influence of recombination energy release on mass
ejection, we repeat our reference run discussed in
Sect.~\ref{sec:reference_run} with an ideal gas \ac{EoS} thus ignoring
ionization effects. We compare the results with our reference run,
that employs the OPAL \ac{EoS} and assumes local thermalization of
recombination energy without cooling losses. Starting out from the
same \mesa model, the relaxation procedure outlined in
Sect.~\ref{sec:relaxation} is repeated, but this time employing an
ideal gas \ac{EoS}. With the thus obtained \ac{RG} star model, a
binary system is set up adopting the same companion mass of $M_2 =
0.08 \, \msol$ and the same initial orbital parameters as in our
reference simulation.

Fig.~\ref{fig:eos} compares the evolution of unbound mass fraction
according to the thermal energy criterion for both simulations.  We
reach a final value of $f_\mathrm{th} = 77.8\%$ in the reference run
accounting for ionization effects, and the value is still increasing
at the end of the simulation. In contrast, very little mass is ejected
when applying the ideal gas \ac{EoS}. Here, only $f_\mathrm{th} =
7.3\%$ are unbound at the end of the simulation and this fraction is
hardly increasing any longer.

This clearly emphasizes the importance of ionization energy and
recombination processes for the ejection of the envelope for the
considered case with $M_2 = 0.08 \, \msol$. Companions of even lower
masses can certainly not eject the envelope when only tapping the
orbital energy reservoir. As mentioned above, recombination energy is
released when the gas of the envelope has sufficiently expanded and
cooled to allow for electron captures.  This is reflected in the fact
that the unbinding only starts when the inspiral is already well
underway. Since the envelope of the progenitor \ac{RG} is only weakly
bound, even small perturbations by low-mass companions cause a
significant release of ionization energy that can ultimately lead to a
nearly complete unbinding.

\begin{figure*}
  \resizebox{\hsize}{!}{\includegraphics{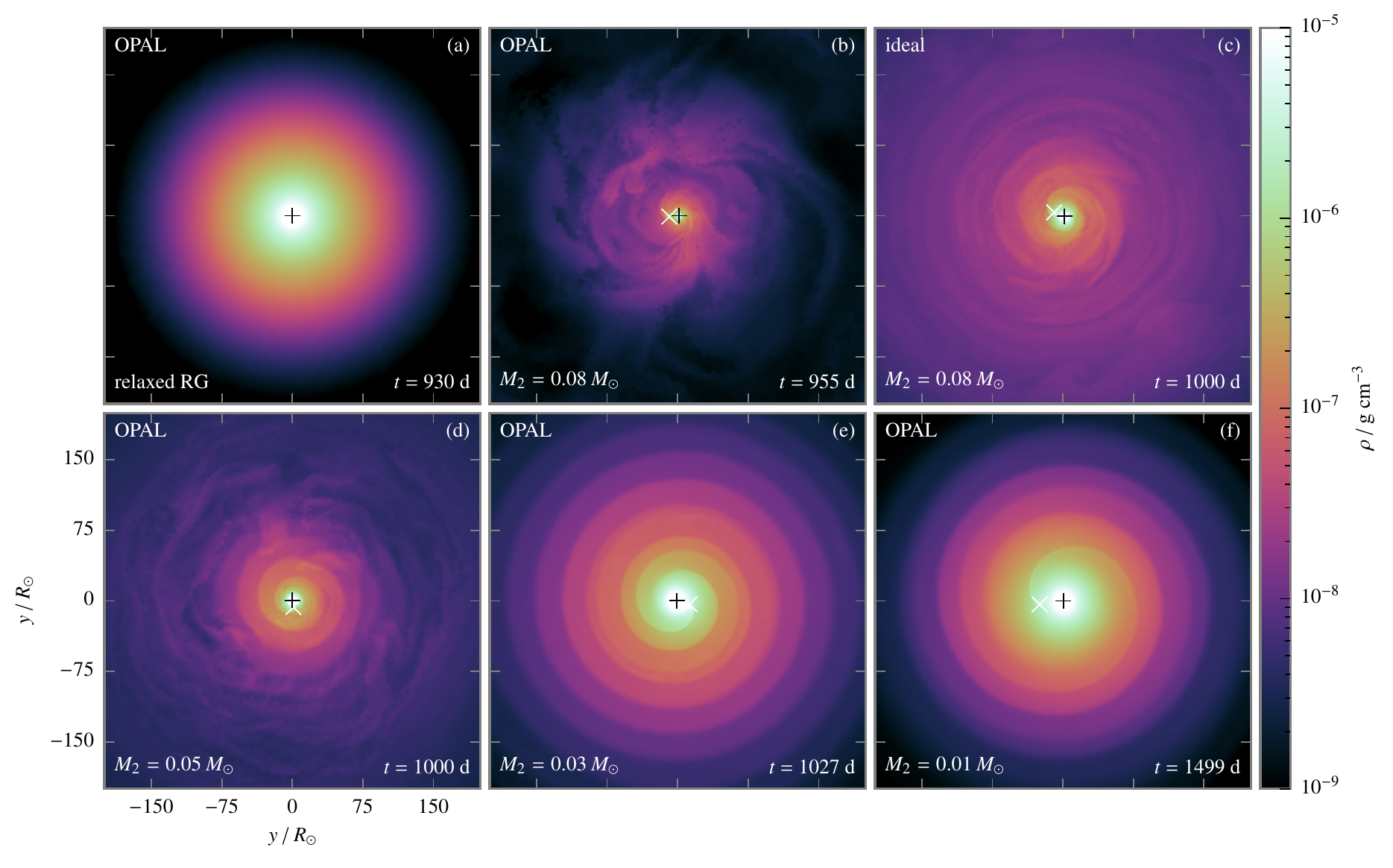}}
  \caption{Density snapshots in the orbital plane for different
    systems. Panel (a) shows the inital state of the primary \ac{RG}
    star before a companion is added. Here, the time indicates that of
    the relaxation run. For the binary simulations in panels (b) to
    (f), the mass of the companion is given on the bottom left side of
    each panel, the time after adding the companion at the bottom
    right and the equation of state at the top left. The positions of
    the cores of the \ac{RG} primary star and the companion are marked
    by an \(\times\) and + respectively. Each frame is centered on the
    center of mass of the binary system.}
  \label{fig:lastSnap}
\end{figure*}

Fig.~\ref{fig:lastSnap} illustrates the final state of the two
simulations. Its Panel (b) shows a density slice through the orbital
plane at the end of the simulation with the OPAL \ac{EoS} where
ionization energy is included. Comparing with the initial \ac{RG}
primary in panel (a) shows that outer envelope is largely lost and the
low density of the remaining material exhibits an irregular pattern.
In contrast, the envelope material of the \ac{RG} for the run without
ionization effects in panel (c) is notably less disrupted and
smoother. The envelope has expanded [cf.\ panel (a)], but since
ionization energy is not taken into account, no recombination energy
can be released and only a small amount of mass is unbound.

\subsection{Varying companion masses}
\label{sec:companion}

\begin{table*}[h]
    \centering
    {\def\arraystretch{1.2}
        \caption{Overview of the simulation runs with different companion masses.} 
        \label{tab:companionmass}
    \begin{tabular}{*{13}{c}} 
        \hline \hline
        {$M_2 / \msol $} & {$q$} & {$t_\mathrm{f} / \mathrm{d}$ } & {$a_\mathrm{i} /
        \rsol$} & {$a_\mathrm{f} / \rsol$} & {$P_{\mathrm{i}} / \mathrm{d} $} & {$P_{\mathrm{f}} /
        \mathrm{d}$} & {$e_\mathrm{f}$} & {$ \Delta E / E_{0} $} & {$\Delta J /
        J_{0} $} & $ f_\mathrm{int} $ & $f_\mathrm{th}$ & EoS  \\ \hline  
        0.08 & 0.10 & 955 & 164.3 & 10.4 & 329.4 & 5.3 & 0.042 & 0.6\% & -1.1\% & 99.7\% & 77.8\% & OPAL \\
        0.05 & 0.06 & 1000 & 154.7 & 7.5 & 309.3 & 3.2 & 0.033 & 0.2\% & -1.4\% & 98.1\% & 52.1\% & OPAL \\
        0.03 & 0.04 & 1027 & 146.2 & 14.2 & 289.9 & 8.6 & 0.062 & 0.3\% & -1.5\% & 71.8\% & 21.8\% & OPAL \\
        0.01 & 0.01 & 1500 & 133.4 & 12.4 & 257.9 & 7.4 & 0.032 & 0.6\% & -0.8\% & 46.3\% & 16.6\% & OPAL \\
        \addlinespace[0.5ex]
        0.08 & 0.10 & 1000 & 163.9 & 11.6 & 328.1 & 6.3 & 0.026 & 0.2\% & -1.7\% & 7.3\% & 7.3\% & ideal \\
        \hline
    \end{tabular}}
\end{table*}

\begin{figure}
  \resizebox{\hsize}{!}{\includegraphics{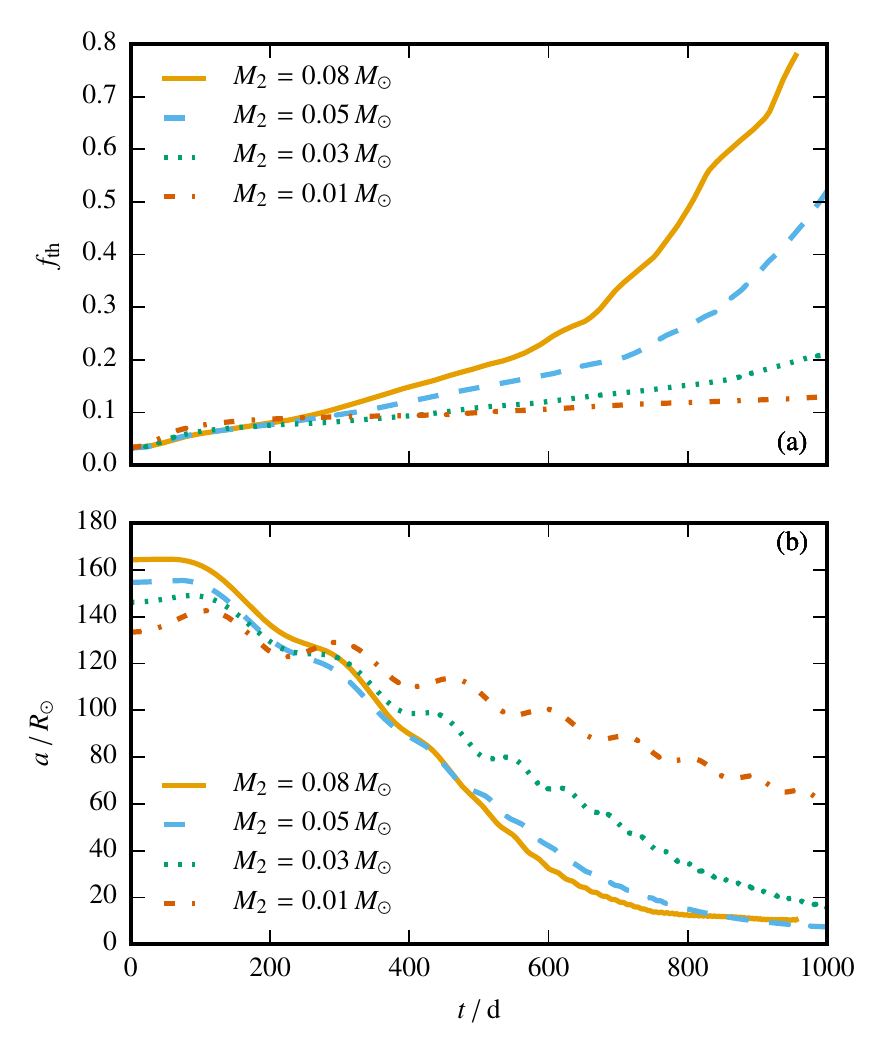}}
  \caption{Fraction of unbound mass $f_\mathrm{th}$ accroding to the
    thermal energy criterion and orbital separation $a$ between the
    core particles over time for simulations with the indicated
    companion masses $M_2$.}
  \label{fig:companion_masses}
\end{figure}

To study the influence of the companion's mass on the envelope
ejection, we conduct further simulations with identical setups as for
the reference run in Sect.~\ref{sec:reference_run}, but with lower
$M_2$ (see Table \ref{tab:companionmass}). With its $0.08 \, \msol$,
companion of the reference simulation marks the limit above which
hydrogen burning ignites and the objects become true stars. With our
parameter study, we map out the mass range of \ac{BD}s down to the
most massive giant planets with $M \lesssim 0.01 \, \msol$.
Table~\ref{tab:companionmass} summarizes parameters of our
simulations. For comparison, we also include our similation with the
ideal gas \ac{EoS}.  In all runs, relative error in energy stays below
1\% and in angular momentum under 2\%. We list the initial separations
between the core particles $a_\mathrm{i}$ and the initial orbital
periods $P_\mathrm{i}$, as well as the corresponding values at the end
of the respective simulations ($a_\mathrm{f}$ and $P_\mathrm{f}$).

The top panel of Fig.~\ref{fig:companion_masses} shows the evolution
of the fraction of unbound mass $f_\mathrm{th}$ according to the
thermal energy criterion. Higher companion masses cause stronger
perturbation of the envelope and lead to increased mass
ejection. However, for masses as low as \(M_2 = 0.05 \, \msol\) (\(q =
0.06\)), 52.1\% of the envelope become unbound under the thermal
energy criterion until the end of our simulation. It is also clear
that material is still being ejected from the system at this point. If
we apply the internal energy criterion which also accounts for
ionization energy, the fraction increases to $f_\mathrm{int} = 98.1\%$
(see Table~\ref{tab:companionmass}). The simulation with the lowest
companion mass, however, shows a different picture (see upper panel of
Fig.~\ref{fig:companion_masses}): the unbound mass according to the
thermal energy criterion quickly reaches about 10\% and then
stagnates.  Only 16.6\% of the envelope mass is unbound at the end of
the simulation. When employing the internal energy criterion, we find
$f_\mathrm{int} = 0.46$, \ie even if all available ionization energy
is used, less than half of the envelope will eventually become
unbound.

In the bottom panel of Fig.~\ref{fig:companion_masses}, the orbital
separation between the core particles over time is plotted.  While for
the runs with the companion masses of \(0.08 \, \msol\) and \(0.05 \,
\msol\) the separation between the core particles shows a
qualitatively similar evolution, we see a distinct behavior for the
companion with the lowest mass of $0.01 \, \msol$, where the orbital
separation slowly but steadily decreases.

These results indicate that around $M_2 = 0.03 \msol$ there is a
companion mass threshold below which the dynamic interaction with the
envelope is not strong enough to trigger significant envelope ejection
(we discuss this further in Sect.~\ref{sec:minimum-mass}).

This qualitatively different behavior can also be seen when comparing
the density slices through the orbital plane at the end of the
different simulations in Fig.~\ref{fig:lastSnap}. In panel (a), the
final state of the \ac{RG} at the end of the relaxation run is given
before placing the companion. As discussed above, a direct comparison
of the relaxed model and the final state after the binary interaction
with the \(M_2 = 0.08 \, \msol\) (\(q = 0.10\)) companion makes the
degree of perturbation of the envelope material apparent. Large scale
instabilities have emerged and the smooth envelope is notably
disrupted. The same can be observed for a companion mass of \(M_2 =
0.05 \, \msol\) (\(q = 0.06\)), which is shown in panel (d). However,
the disruption in the inner part is not as strong and more material
remains in the outer regions. This is not surprising considering the
slower mass ejection in this case (see top panel of
Fig.~\ref{fig:companion_masses}). As the unbound mass fraction
$f_\mathrm{th}$ is still increasing and $f_\mathrm{int} \approx 1$,
this marks only a snapshot in an evolution that ultimately will lead
to almost complete envelope removal.  This is different for the two
lower companion masses. Their density slices in panels (e) and (f) of
Fig.~\ref{fig:companion_masses} display a less perturbed
envelope. While layered spiral structures have emerged, no large scale
perturbations occur.  For the companion mass of \(M_2 = 0.03 \,
\msol\) (\(q = 0.04\)), the radius of the star appears to have
slightly expanded, but for \(M_2 = 0.01 \, \msol\) (\(q = 0.01\)) the
expansion is marginal if present at all (see
Sect.~\ref{sec:minimum-mass}).

\subsection{Minimum mass for envelope ejection}\label{sec:minimum-mass}

\begin{figure*}
  \resizebox{\hsize}{!}{\includegraphics{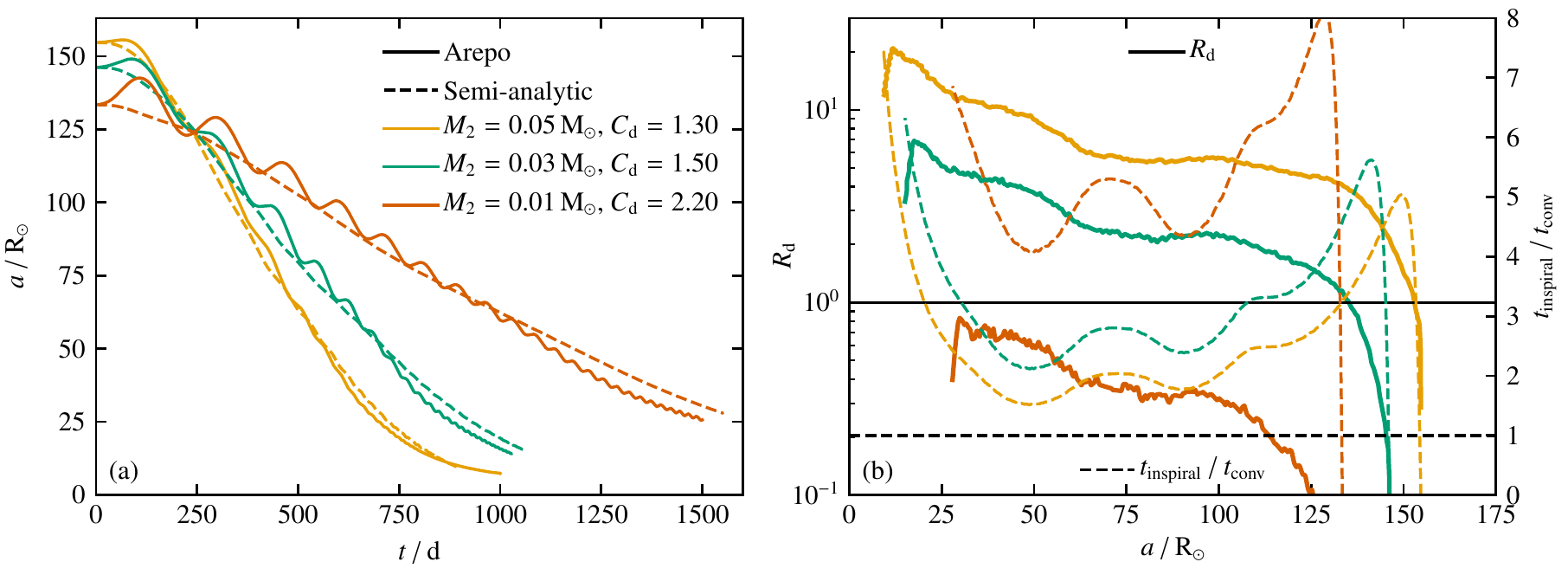}}
  \caption{Comparison of the evolution of the orbital separation of
    the various \arepo runs with the simplified drag model (left panel
    a), and the dynamic response number $R_\mathrm{d}$ and the local
    ratios of inspiral and convective turn-over time (right panel
    b). The evolution of the orbital separation of a $0.08 \, \msol$
    companion is similar to that of the $0.05 \, \msol$ companion
    (Fig.~\ref{fig:companion_masses}) and therefore not shown here for
    clarity.}
  \label{fig:minimum-mass}
\end{figure*}

We have seen from our simulations that lower-mass companions eject
less mass (Table~\ref{tab:companionmass} and
Fig.~\ref{fig:companion_masses}). Even more so, the runs with
companion masses of $0.03 \, \msol$ and $0.01 \, \msol$ have envelope
ejection fractions of $f_\mathrm{th}\lesssim 20\%$ over the course of
our computations. This suggests that there exist a minimum companion
mass below which the envelope of the \ac{RGB} star is not perturbed
strongly enough to cause significant envelope ejection. To further
illuminate this lower mass threshold qualitatively, we consider a
companion of mass $M_\mathrm{2}$ orbiting in the unperturbed envelope
of our \ac{RGB} star under the influence of a drag force \citep[see
  also][]{macleod2015a, macleod2017b, chamandy2019a},
\begin{align}
    F_\mathrm{drag} = \dot{M} v_\mathrm{rel} = \pi R_\mathrm{a}^2 \rho
    ( v_\mathrm{rel}^2 + c_\mathrm{s}^2 )^{1/2} v_\mathrm{rel}.
    \label{eq:drag-force-1}
\end{align}
In this equation, $v_\mathrm{rel}$ is the relative velocity of the
companion with respect to the bulk rotational velocity of the \ac{RGB}
star's envelope, $\rho$ and $c_\mathrm{s}$ are the local density and
sound speed in the unperturbed \ac{RGB} star, respectively, and
$R_\mathrm{a}$ a mass-accretion radius that we approximate by
Bondi--Hoyle--Littleton theory \citep{bondi1944a, bondi1952a},
\begin{align}
    R_\mathrm{a} = \frac{2 G M_2}{v_\mathrm{rel}^2 + c_\mathrm{s}^2}.
    \label{eq:bondi-radius}
\end{align}
We further multiply the drag force by a drag coefficient $C_\mathrm{d}$ such
that we finally have
\begin{align}
    F_\mathrm{drag} = C_\mathrm{d} \frac{4\pi G^2 M_2^2 \rho
      v_\mathrm{rel}}{( v_\mathrm{rel}^2 + c_\mathrm{s}^2 )^{3/2}}.
    \label{eq:drag-force}
\end{align}
We initially place the companion as in the \arepo runs (\ie at the
same initial orbital separation with an initial velocity according to
a Keplerian orbit) and assume that the \ac{RGB} star's envelope
rotates as a solid body that is in 95\% co-rotation with the
companion's initial orbit. We use the unperturbed envelope structure
from the \mesa model of the \ac{RGB} star and then solve the equation
of motion of the companion under the influence of the drag force in
Eq.~(\ref{eq:drag-force}) with a fourth-order Runge--Kutta method. In
this simplified model, the dynamical response of the envelope to the
deposition of released orbital energy is neglected. The drag
coefficient $C_\mathrm{d}$ is adjusted by hand such that the temporal
evolution of the orbital separation is close to that found in the
\arepo simulations (Fig.~\ref{fig:minimum-mass}a).

The instantaneous energy injection into the \ac{RGB} star's envelope
within this simplified model is $\dot{E}_\mathrm{drag} =
F_\mathrm{drag} v_\mathrm{rel}$.  With this energy injection rate, we
define a local inspiral time,
\begin{align}
    t_\mathrm{inspiral} = \frac{\Delta
      E_\mathrm{orb}}{\dot{E}_\mathrm{drag}},
    \label{eq:inspiral-time}
\end{align}
where $\Delta E_\mathrm{orb}=G M_1 M_2 / 2 (1/a_t - 1/a_i)$ is the
absolute change of the orbital energy from the beginning to the
current time (with $a_t$ the current and $a_i$ the initial orbital
separation). In the following, we compare this inspiral time to the
convective turn-over time,
\begin{align}
    t_\mathrm{conv} = \frac{\alpha_\mathrm{mlt} H_P}{v_\mathrm{conv}},
\end{align}
where $H_P=P/(g\rho)$ is the local pressure scale height with pressure
$P$, density $\rho$, and gravitational acceleration $g$,
$v_\mathrm{conv}$ is the velocity of convective eddies and
$\alpha_\mathrm{mlt}$ relates to the convective mixing efficiency
within mixing-length theory ($\alpha_\mathrm{mlt}=2.0$ in our
model). This is the relevant time scale in our problem, because
convection dominates the energy transport in the envelope of the
\ac{RGB} star (photon diffusion only plays a role in the outermost
envelope and the inspiral time is always faster than the photon
diffusion time for separations $a\lesssim\SI{140}{\rsol}$).

This is of course only true if convection is still established despite
a companion star perturbing the envelope by its inspiral. In the
presence of rotational fluid motions, the Solberg--H{\o}iland
criterion can be used to assess convective stability
\citep[\eg][]{kippenhahn2012a}. \citet{ohlmann2016a} find in their
\ac{CE} simulation that convection is partly suppressed in the
stirred-up envelope. Also in outflowing envelope material that is
unbound and hence no longer in hydrostatic equilibrium, convection
cannot occur. The following discussion is therefore only true if
convection can still contribute to the energy transport in the
perturbed \ac{RGB} star's envelope \citep[see also][]{wilson2020a}.

We further define a local dynamic response number in the \ac{RGB} star's
envelope,
\begin{align}
    R_\mathrm{d} = \frac{\dot{E}_\mathrm{drag} t_\mathrm{conv}}{E_\mathrm{int}},
    \label{eq:dyn-response-number}
\end{align}
that compares the local energy injection over a convective turn-over
time to the local internal energy of the gas in the envelope. The so
defined dynamic response number is a measure to judge whether the
envelope is expected to react dynamically, potentially leading to mass
ejection. Energy, which is injected locally in the envelope, may be
transported to the stellar surface by convection, where it can be
radiated away. To trigger a dynamic response of the convective
envelope, one has to
\begin{enumerate}[label=\textnormal{(\roman*)}]
  \item inject energy faster than can be transported to the surface by
  convection and
  \item inject more energy than the local binding or equivalently internal
  energy of the gas (the unperturbed envelope is in virial equilibrium).
\end{enumerate}
For $R_\mathrm{d}\gg1$, the energy injection into the \ac{RGB} star's
envelope over a convective convective turn-over time because of the
drag force acting on the companion is much larger than the local
internal energy of the gas.  Therefore, a dynamical response of the
envelope is expected and this will likely lead to significant mass
ejection. Conversely, for $R_\mathrm{d}\lesssim 1$, the injected
energy may simply be transported to the stellar surface by convection
where it is lost by radiation. In this case, the envelope will not
dynamically respond to orbital energy release and little or no
envelope ejection is expected.

In Fig.~\ref{fig:minimum-mass}b, we plot the dynamic response number
and the ratio of the inspiral and convective turn-over timescale for
companion masses of $0.05 \, \msol$, $0.03 \, \msol$, and $0.01 \,
\msol$. The inspiral time is longer than the convective turn-over time
in all cases, implying that convective energy transport is indeed
relevant. Because $F_\mathrm{drag} \propto {M_2}^2$, the inspiral time
is longer for less massive companions and convection becomes more
important in transporting away the locally injected energy. Moreover,
the total injected energy, \ie essentially the available orbital
energy $E_\mathrm{orb}\propto M_2$, is also smaller for less massive
companions. These two aspects are reflected in the dynamic response
number $R_\mathrm{d}$ in Fig.~\ref{fig:minimum-mass}b: $R_\mathrm{d}$
is of order 5--10 for the $0.05 \, \msol$ companion, but only of the
order of a few for $0.03 \, \msol$ and lower than $1$ for $0.01 \,
\msol$. This implies that a dynamic response and hence significant
envelope ejection is expected for our ${>}\,0.05 \, \msol$ companions,
while this is no longer the case for a $0.01 \, \msol$ companion.  Our
$0.03 \, \msol$ companion setup marks a marginal case.

\begin{table}
  \centering
  {\def\arraystretch{1.2}
      \caption{Drag coefficient $C_\mathrm{d}$, tidal \& evaporation radii 
      ($r_\mathrm{tid}$, $r_\mathrm{eva}$), and corresponding mass coordinates 
      ($m_\mathrm{tid}$, $m_\mathrm{eva}$) from the semi-analytic models for 
      the different companion masses $M_2$.} 
      \label{tab:semi-analytic}
  \begin{tabular}{*{6}{c}} 
      \hline \hline
      {$M_2 / \msol $} & {$C_\mathrm{d}$} & {$r_\mathrm{tid} / \rsol$} &
      {$m_\mathrm{tid} / \msol$} & {$r_\mathrm{eva} / \rsol$} & {$m_\mathrm{eva}
      / \msol$}  \\ 
      \hline  
      0.08 & 0.95 & 0.04 & 0.4616 & 0.11 & 0.4622 \\
      0.05 & 1.30 & 0.05 & 0.4617 & 0.17 & 0.4625 \\
      0.03 & 1.50 & 0.05 & 0.4617 & 0.31 & 0.4628 \\
      0.01 & 2.20 & 0.06 & 0.4619 & 1.27 & 0.4638 \\
      \hline
  \end{tabular}}
\end{table}

During the \ac{CE} phase, the inspiralling companion may be tidally
disrupted and could evaporate. We compute the tidal disruption radius
as $r_\mathrm{tid}=R_2 (\rho/\rho_2)^{1/3}$, where $R_2=0.1 \, \rsol$
is the characteristic radius \citep{chabrier2000a} and $\rho_2$ is the
average density of an inspiralling \ac{BD} companion. The evaporation
radius $r_\mathrm{eva}$ is defined as that radius where the sound
speed of the ambient \ac{RGB} star's envelope equals the escape
velocity from the inspiralling companion (see \eg \citealt{livio1984a,
  soker1998a, nelemans1998a}, but see also \citealt{jia2018a} for an
alternative criterion). The tidal disruption radius is always smaller
than the radius at which evaporation is expected to become important
(Table~\ref{tab:semi-analytic}). The evaporation radius is larger in
lower-mass companions and it is $1.3 \, \rsol$ for $M_2 = 0.01 \,
\msol$. In case there is no envelope ejection, we expect the companion
to start evaporating before it may finally be tidally disrupted. Both,
tidal disruption and evaporation would take place near the very bottom
of the convective envelope at mass coordinates of $0.462$ to $0.464 \,
\msol$ such that the companion's material is likely mixed throughout
the convective envelope of the \ac{RGB} star.

In summary, we find that for a number of arguments there is a lower
mass threshold below which \ac{CE} ejection becomes
impossible. Contrary to the estimates of \citet{soker1998a} and
\citet{nelemans1998a}, however, this threshold is not determined by
the evaporation of the companion, although this may also be its fate
in our models (see Sect~\ref{sec:final-fate}). It is rather set by the
ability of the companion to trigger a dynamical response of the
envelope. In two different approaches -- \ac{3D} hydrodynamic
simulations and \ac{1D} semi-analytical modeling -- we have explored
the value of the threshold. It is important to note that both
treatments capture somewhat different parts of the relevant physical
processes: While our \ac{3D} hydrodynamical simulations account for
the expansion of the envelope material because of the release of
orbital energy, which is neglected in the semi-analytical treatment,
they most likely do not fully resolve convection in the envelope. Both
effects, however, are essential to determine the mass threshold,
because for a successful envelope ejection, the above discussed
conditions (i) and (ii) have to be met. On the one hand, our \ac{3D}
hydrodynamic simulations determine the correct energy needed locally
for mass unbinding, but it remains unclear whether released energy
leads to a dynamic envelope expansion and mass ejection rather than
being transported away by convection. The spiral structure seen in the
simulations (see Figs.~\ref{fig:time_series_rho} and
\ref{fig:lastSnap}), however, question the persistence of initial
global convective motions in this phase. On the other hand, our
semi-analytic models fail to correctly determine the local binding
energy of envelope material because it should have expanded in the
inspiralling process -- at least in the cases of more massive
companions. We argue that in the case of low-mass companion, expansion
will be inefficient and therefore our model (although not correctly
describing \ac{CEE} high companion masses) still provides a meaningful
estimate of the mass threshold for envelope unbinding. The fact that
both models predict the threshold to be somewhere around $0.03\,
\msol$ is reassuring and consequently this value marks our current
best estimate for the mass threshold above which one can expect
envelope ejection in a \ac{CE} phase with our \ac{RGB} star.

\section{Discussion}\label{sec:discussion}

\subsection{Final fate}\label{sec:final-fate}

For companions of $M_2\gtrsim0.03\,\msol$, the inspiral results in a
strong dynamical response of the \ac{RGB} star and we expect mass
ejection.  For $M_2\gtrsim0.05\,\msol$, almost the whole envelope
(98.1\%, Table~\ref{tab:companionmass}) may be ejected if all of the
available ionisation energy at the end of the run can be employed in
unbinding the envelope. For $M_2\approx0.03\text{--}0.05\,\msol$, it
seems that not the whole envelope can be ejected (e.g.\ we find
$f_\mathrm{int}=71.8\%$ for $M_2=0.03\,\msol$). The envelope ejection
fractions are still increasing at the end of our simulations
(Fig.~\ref{fig:companion_masses}), so the above companion mass ranges
for which full and partial envelope ejection are expected will likely
be shifted to lower values.

In case of full envelope ejection ($M_2\gtrsim0.05\,\msol$), our final
orbital separations appear to have converged. However, this is not to
say that the orbit will not change anymore. For example, there is
still matter inside the binary orbit at the end of our simulation that
may affect the final orbital configuration. Also, if some high
specific-angular-momentum material of the former envelope remains
bound to the binary system, a circumbinary disk may form that could
exert a torque on the inner binary and thereby lead to further orbital
shrinkage and possibly eccentricity pumping
\citep[\eg][]{artymowicz1991a, artymowicz1994a, dermine2013a,
  reichardt2019a}.  If such a disk is massive enough and long lived,
there could even be planet formation as has been suggested in the
literature in other and similar situations
\citep[\eg][]{podsiadlowski1993a, perets2010a, beuermann2010a,
  volschow2014a}.

For partial envelope ejection ($M_2\approx0.03\text{--}0.05\,\msol$), the
\ac{RGB} star will retain parts of its original envelope and other, high
specific-angular-momentum parts may settle into a circumstellar disk. Retaining
only a few percent of the original envelope mass is usually enough to maintain
giant-star-like radii. As can be seen in our simulations, the dynamical drag on
the companion is small at the end of our computations such that the orbital
separation has settled to a final value (Fig.~\ref{fig:companion_masses}). From
here on, the spiral-in slows down and we anticipate that the system enters a
self-regulated phase where the released orbital energy may be transported to the
stellar surface by turbulent convection and radiated away \citep[see
also][]{meyer1979a, podsiadlowski2001a}. The companion will then likely dissolve
inside the envelope by evaporation rather than being tidally disrupted
(Table~\ref{tab:semi-analytic}). During the self-regulated \ac{CE} phase, energy
is continuously injected into the envelope of the \ac{RGB} star, which can
trigger pulsations and thereby help to eject the further envelope material
\citep{clayton2017a}. Ultimately, this may lead to the full ejection of the
envelope.

In case of no ($M_2\lesssim0.03\,\msol$) and partial envelope ejection, the
further evolution of the \ac{RGB} star may be affected in various ways as
studied by several authors and groups \citep[see \eg][and references
therein]{soker1998a, siess1999a, israelian2001a, stephan2020a}:
\begin{enumerate}[label=\textnormal{(\roman*)}]
  \item Orbital energy is injected into the \ac{RGB} star and this additional
  energy will subsequently be radiated away during a phase in which the star
  regains thermal equilibrium. Such a transient will roughly last for a thermal
  timescale of that part of the envelope in which energy has been injected. 
  \item Some fraction of the initial orbital angular momentum has been ingested
  into the \ac{RGB} star such that it may rotate rapidly.
  \item In case of partial envelope ejection, the resulting star has an unusual
  core-envelope structure for its evolutionary stage. During core-helium burning
  as a horizontal-branch star, this peculiar structure may introduce features to
  the horizontal-branch morphology of stellar populations \citep[see
  also][]{dcruz1996a}.
  \item As discussed in Sect.~\ref{sec:minimum-mass}, the companion likely
  evaporates inside the convective envelope such that its chemical constituents
  are mixed up to the surface of the \ac{RGB} star. Moreover, the dissolution of
  brown dwarfs and planets may activate hot-bottom burning during which lithium
  can be produced and subsequently mixed to the stellar surface. Altogether,
  planet-eating stars may show unusual surface abundances. 
  \item The rapid rotation of the convective envelope may boost a magnetic
  dynamo such that the star obtains a strong magnetic field and might show
  enhanced magnetic activity. Furthermore, during the \ac{CEE} as well as
  main-sequence mergers, the magneto-rotational instability is found to amplify
  an initially weak seed magnetic field and thereby highly magnetizes the
  stellar envelope \citep[][]{ohlmann2016b, schneider2019a}. A dynamo operating
  in this convective envelope could then lead to an even stronger magnetic field
  than what may be expected otherwise from a convective dynamo in a rapidly
  rotating envelope \citep[\eg][]{featherstone2009a, braithwaite2017a}. As
  mentioned above, our \ac{3D} magnetohydrodynamic simulations allow us to
  follow such effects and we will discuss them in a forthcoming publication.
\end{enumerate}

\subsection{Comparison to observations of sdB stars}\label{sec:comp-observations}

Originally, our study is motivated by the strong evidence for low-mass
companions of \ac{sdB} stars found in observations. More specifically,
the analysis of HW\,Vir systems shows that a significant fraction of
\ac{sdB}s are orbited by close companions in the mass regime of
\ac{BD}s \citep{schaffenroth2018a}. \citet{geier2011a} determine the
companion mass of SDSS\,J082053.53+000843.4 to be between $0.045$ and
$0.068\, \msol$.  Moreover, SDSS\,J162256.66+473051.1 and V2008$-$1753
likely harbour \ac{BD} companions with masses of $0.064\, \msol$ and
$0.069\, \msol$, respectively \citep{schaffenroth2014a,
  schaffenroth2015a}. The mass of the close companion to AA\,Dor
($0.079\, \msol$) is very close to the hydrogen-burning limit and it
might therefore also be a massive \ac{BD}
\cite{vuckovic2016a}. Furthermore, \citet{schaffenroth2014a} report
the discovery of two non-eclipsing sdB binaries with very small
minimum masses for the companions. Both the cool companions of
PHL\,457 ($>0.027\, \msol$) and CPD$-$64$^{\circ}$481 ($>0.048\,
\msol$) might be \ac{BD}s. SdBs with \ac{BD}s will eventually evolve
to detached white dwarf systems with \ac{BD} companions. Of such
systems nine are known, seven of them with white dwarfs very close to
the masses expected for \ac{sdB} stars. Their companion masses fall
into the range of $0.05$ to $0.07\, \msol$ \citep[][and references
  therein]{casewell2018a}.  These systems might therefore have formed
from the scenario we study in this work.

Our simulations show successful envelope ejection for systems with
\ac{BD} companions as indicated by the observations. Also, the lower
mass limit determined here is supported by observations: Despite
considerable effort no giant planet has so far been identified in
close orbit around a hot subdwarf yet \citep{schaffenroth2019a,
  casewell2018a}. However, there may be a bias because the lowest-mass
companions are expected at longer periods of about
$\SI{0.3}{\day}$. Otherwise, the companion would exceed its Roche lobe
leading to mass transfer onto the primary \citep[see Fig. 14
  of][]{schaffenroth2019a}. Such long-period systems are harder to
follow-up and only few of them have been investigated until now.

Despite the success of our simulations to reproduce the formation
scenario inferred from observations, the resulting systems do not
match the observed orbital separations of about $0.4 \, \rsol$ to $1.3
\, \rsol$ (see \citealp{casewell2018a}, and references therein;
\citealp{schaffenroth2018a}). Our simulations yield final orbital
separations between the cores that are larger by a factor of
$\sim$$10$. This points to physical processes impacting the final
separations that are not accounted for in our simulations, such as
energy loss by convection \citep{wilson2020a} or mass loss of the
companion by continuous evaporation during the inspiral. It is also
possible that the resulting systems forms a circumbinary disk on
timescales longer than those followed in our simulations, that
interacts with the inner binary changing its orbital
parameters. Another possibility is that higher-mass \ac{RG} primaries
lead to a deeper spiral-in of the companion. Whether this can account
for the observed difference has to be tested in simulations.

\section{Conclusion} \label{sec:conclusion}

In numerical and semi-analytical approaches, we address the question
under which conditions \ac{sdB} stars can form from stars at the tip
of the red giant branch when interacting with low-mass
companions. Observations of eclipsing close binary systems consisting
of \ac{sdB} and cool low-mass companions \citep[HW\,Vir type systems,
  see][and references therein]{schaffenroth2019a} indicate that such a
formation channel is indeed realized in nature.

Based on three-dimensional hydrodynamic simulations of common-envelope
evolution, we show that envelope ejection in such systems is possible
even with companions in the mass range of \ac{BD}s -- provided that
the ionization energy released when the envelope expands is
thermalized locally and supports gravitational unbinding of the
material. This limit is tested in our simulations as well as a model
where no ionization energy is used, in which case little envelope
material is expelled. The question of whether recombination energy can
significantly increase mass ejection compared to cases where mass loss
is powered by the release of orbital energy only has been discussed
controversially by \citet{grichener2018a}, but see
\citet{ivanova2018a}. They, however, refer to systems that differ from
those under consideration here. Recent \ac{3D} hydrodynamic
simulations of \ac{CEE} with asymptotic-giant branch primary stars
(Sand et al., in prep.) suggest that in this case the majority of
recombination energy is indeed released in optically thick regions and
contributes to envelope removal. Although this ultimately has to be
confirmed in simulations accounting for radiative transport, the
situation is even more favorable for the systems considered here:
Low-mass companions lead to less efficient expansion of the envelope
so that if the necessary conditions are reached for recombination, it
will be in rather dense and optically thick regions. This strongly
suggests that trapping of released ionization energy is a realistic
assumption.

Determining the lowest companion mass admissible for successful
envelope ejection is a fundamental problem of common-envelope
evolution modeling. A safe upper limit is that sufficient orbital
energy is deposited in the envelope to overcome its binding
energy. Hydrodynamical simulations show that this is usually prevented
by envelope expansion inhibiting energy transfer by tidal drag before
significant amounts of material are expelled. We confirm this for our
specific setup. The envelope expansion, however, leads to the release
of ionization energy that can further unbind envelope
material. Therefore, we suggest as a lower limit for successful
envelope ejection a companion mass that triggers a dynamical response
of the stellar envelope. Combining the findings of our \ac{3D}
hydrodynamic \ac{CE} simulations with a semi-analytic model, we
conclude that in our setup this is the case for companions with more
than about $0.03\, \msol$. This suggests that for giant planets as
companions the formation of \ac{sdB} stars in common envelope episodes
with stars at the tip of the red giant branch is difficult to
achieve. Our result is consistent with currently available
observations of \ac{sdB} stars with low-mass companions.  A failure to
eject the envelope may lead to recurrent CE episodes eventually
removing the envelope, the formation of a circumbinary disk, or an
evaporation of the companion inside the envelope with several
implications for observables of the remaining \ac{RGB} star.

The exact threshold, however, is likely to also depend on the mass of
the primary star: for higher primary masses, more envelope material
needs to be lifted, which may require a larger companion mass. This
has to be explored in future simulations. Another effect that has not
been accounted for in our models but may alter the threshold is a
potential mass gain or loss of the companion due to accretion or
ablation when spiralling into the envelope. Although we find that
complete evaporation is not the key to determine the lowest companion
mass for successful envelope ejection, it may be a continuous process
that changes the companion mass in the evolution.

The final orbital separation remains an open question. Compared with
observations, our simulations predict values that are larger by a
factor of about 10. This may imply that the observed systems had more
massive \ac{RG} primaries than assumed in our simulation. It could,
however, also imply that physical processes are important for
determining the orbital parameters of the resulting system which our
current simulations do not account for.

\begin{acknowledgements}
The work of MK, FS, and FKR is supported by the Klaus Tschira
Foundation. SG and VS are partly supported by the Deutsche
Forschungsgemeinschaft (DFG) through grant GE 2506/9-1
\end{acknowledgements}

\begin{acronym}
    \acro{1D}{one-dimensional}
    \acro{3D}{three-dimensional}
    \acro{BD}{brown dwarf}
    \acro{CEE}{common envelope evolution}
    \acro{EoS}{equation of state}
    \acro{RGB}{red giant branch}
    \acro{RL}{Roche lobe}
    \acro{sdB}[sdB]{hot subluminous B-type}
    \acrodefplural{sdB}[sdB]{Hot subluminous B-type} 
    \acro{CE}{common envelope}
    \acro{RG}{red giant}
    \acro{MS}{main sequence}
    \acro{HSE}{hydrostatic equilibrium}
\end{acronym}


\begin{thebibliography}{57}
\expandafter\ifx\csname natexlab\endcsname\relax\def\natexlab#1{#1}\fi

\bibitem[{{Artymowicz} {et~al.}(1991){Artymowicz}, {Clarke}, {Lubow}, \&
  {Pringle}}]{artymowicz1991a}
{Artymowicz}, P., {Clarke}, C.~J., {Lubow}, S.~H., \& {Pringle}, J.~E. 1991,
  \apjl, 370, L35

\bibitem[{{Artymowicz} \& {Lubow}(1994)}]{artymowicz1994a}
{Artymowicz}, P. \& {Lubow}, S.~H. 1994, \apj, 421, 651

\bibitem[{{Beuermann} {et~al.}(2010){Beuermann}, {Hessman}, {Dreizler},
  {Marsh}, {Parsons}, {Winget}, {Miller}, {Schreiber}, {Kley}, {Dhillon},
  {Littlefair}, {Copperwheat}, \& {Hermes}}]{beuermann2010a}
{Beuermann}, K., {Hessman}, F.~V., {Dreizler}, S., {et~al.} 2010, \aap, 521,
  L60

\bibitem[{{Bondi}(1952)}]{bondi1952a}
{Bondi}, H. 1952, \mnras, 112, 195

\bibitem[{{Bondi} \& {Hoyle}(1944)}]{bondi1944a}
{Bondi}, H. \& {Hoyle}, F. 1944, \mnras, 104, 273

\bibitem[{{Braithwaite} \& {Spruit}(2017)}]{braithwaite2017a}
{Braithwaite}, J. \& {Spruit}, H.~C. 2017, Royal Society Open Science, 4,
  160271

\bibitem[{{Casewell} {et~al.}(2018){Casewell}, {Braker}, {Parsons}, {Hermes},
  {Burleigh}, {Belardi}, {Chaushev}, {Finch}, {Roy}, {Littlefair}, {Goad}, \&
  {Dennihy}}]{casewell2018a}
{Casewell}, S.~L., {Braker}, I.~P., {Parsons}, S.~G., {et~al.} 2018, \mnras,
  476, 1405

\bibitem[{{Chabrier} \& {Baraffe}(2000)}]{chabrier2000a}
{Chabrier}, G. \& {Baraffe}, I. 2000, \araa, 38, 337

\bibitem[{{Chamandy} {et~al.}(2019){Chamandy}, {Blackman}, {Frank},
  {Carroll-Nellenback}, {Zou}, \& {Tu}}]{chamandy2019a}
{Chamandy}, L., {Blackman}, E.~G., {Frank}, A., {et~al.} 2019, \mnras, 490,
  3727

\bibitem[{{Clayton} {et~al.}(2017){Clayton}, {Podsiadlowski}, {Ivanova}, \&
  {Justham}}]{clayton2017a}
{Clayton}, M., {Podsiadlowski}, P., {Ivanova}, N., \& {Justham}, S. 2017,
  \mnras, 470, 1788

\bibitem[{{D'Cruz} {et~al.}(1996){D'Cruz}, {Dorman}, {Rood}, \&
  {O'Connell}}]{dcruz1996a}
{D'Cruz}, N.~L., {Dorman}, B., {Rood}, R.~T., \& {O'Connell}, R.~W. 1996, \apj,
  466, 359

\bibitem[{{Dermine} {et~al.}(2013){Dermine}, {Izzard}, {Jorissen}, \& {Van
  Winckel}}]{dermine2013a}
{Dermine}, T., {Izzard}, R.~G., {Jorissen}, A., \& {Van Winckel}, H. 2013,
  \aap, 551, A50

\bibitem[{{Duch{\^e}ne} \& {Kraus}(2013)}]{duchene2013a}
{Duch{\^e}ne}, G. \& {Kraus}, A. 2013, \araa, 51, 269

\bibitem[{{Featherstone} {et~al.}(2009){Featherstone}, {Browning}, {Brun}, \&
  {Toomre}}]{featherstone2009a}
{Featherstone}, N.~A., {Browning}, M.~K., {Brun}, A.~S., \& {Toomre}, J. 2009,
  \apj, 705, 1000

\bibitem[{{Geier} {et~al.}(2011){Geier}, {Classen}, \& {Heber}}]{geier2011a}
{Geier}, S., {Classen}, L., \& {Heber}, U. 2011, \apjl, 733, L13

\bibitem[{{Grichener} {et~al.}(2018){Grichener}, {Sabach}, \&
  {Soker}}]{grichener2018a}
{Grichener}, A., {Sabach}, E., \& {Soker}, N. 2018, \mnras, 478, 1818

\bibitem[{{Heber}(1986)}]{heber1986a}
{Heber}, U. 1986, \aap, 155, 33

\bibitem[{{Israelian} {et~al.}(2001){Israelian}, {Santos}, {Mayor}, \&
  {Rebolo}}]{israelian2001a}
{Israelian}, G., {Santos}, N.~C., {Mayor}, M., \& {Rebolo}, R. 2001, \nat, 411,
  163

\bibitem[{{Ivanova}(2018)}]{ivanova2018a}
{Ivanova}, N. 2018, \apj, 858, L24

\bibitem[{{Ivanova} {et~al.}(2013){Ivanova}, {Justham}, {Chen}, {De Marco},
  {Fryer}, {Gaburov}, {Ge}, {Glebbeek}, {Han}, {Li}, {Lu}, {Marsh},
  {Podsiadlowski}, {Potter}, {Soker}, {Taam}, {Tauris}, {van den Heuvel}, \&
  {Webbink}}]{ivanova2013a}
{Ivanova}, N., {Justham}, S., {Chen}, X., {et~al.} 2013, \aapr, 21, 59

\bibitem[{{Jia} \& {Spruit}(2018)}]{jia2018a}
{Jia}, S. \& {Spruit}, H.~C. 2018, \apj, 864, 169

\bibitem[{{Kippenhahn} {et~al.}(2012){Kippenhahn}, {Weigert}, \&
  {Weiss}}]{kippenhahn2012a}
{Kippenhahn}, R., {Weigert}, A., \& {Weiss}, A. 2012, {Stellar Structure and
  Evolution} (Berlin Heidelberg: Springer-Verlag)

\bibitem[{{Livio} \& {Soker}(1984)}]{livio1984a}
{Livio}, M. \& {Soker}, N. 1984, \mnras, 208, 763

\bibitem[{{MacLeod} {et~al.}(2017){MacLeod}, {Antoni}, {Murguia-Berthier},
  {Macias}, \& {Ramirez-Ruiz}}]{macleod2017b}
{MacLeod}, M., {Antoni}, A., {Murguia-Berthier}, A., {Macias}, P., \&
  {Ramirez-Ruiz}, E. 2017, \apj, 838, 56

\bibitem[{{MacLeod} \& {Ramirez-Ruiz}(2015)}]{macleod2015a}
{MacLeod}, M. \& {Ramirez-Ruiz}, E. 2015, \apj, 803, 41

\bibitem[{{Maxted} {et~al.}(2001){Maxted}, {Heber}, {Marsh}, \&
  {North}}]{maxted2001a}
{Maxted}, P.~F.~L., {Heber}, U., {Marsh}, T.~R., \& {North}, R.~C. 2001,
  \mnras, 326, 1391

\bibitem[{{Meyer} \& {Meyer-Hofmeister}(1979)}]{meyer1979a}
{Meyer}, F. \& {Meyer-Hofmeister}, E. 1979, \aap, 78, 167

\bibitem[{{Moe} \& {Di Stefano}(2017)}]{moe2017a}
{Moe}, M. \& {Di Stefano}, R. 2017, \apjs, 230, 15

\bibitem[{{Nandez} {et~al.}(2015){Nandez}, {Ivanova}, \&
  {Lombardi}}]{nandez2015a}
{Nandez}, J.~L.~A., {Ivanova}, N., \& {Lombardi}, J.~C. 2015, \mnras, 450, L39

\bibitem[{{Nelemans} \& {Tauris}(1998)}]{nelemans1998a}
{Nelemans}, G. \& {Tauris}, T.~M. 1998, \aap, 335, L85

\bibitem[{{Ohlmann} {et~al.}(2016{\natexlab{a}}){Ohlmann}, {R{\"o}pke},
  {Pakmor}, \& {Springel}}]{ohlmann2016a}
{Ohlmann}, S.~T., {R{\"o}pke}, F.~K., {Pakmor}, R., \& {Springel}, V.
  2016{\natexlab{a}}, \apjl, 816, L9

\bibitem[{{Ohlmann} {et~al.}(2017){Ohlmann}, {R\"{o}pke}, {Pakmor}, \&
  {Springel}}]{ohlmann2017a}
{Ohlmann}, S.~T., {R\"{o}pke}, F.~K., {Pakmor}, R., \& {Springel}, V. 2017,
  \aap, 599, A5

\bibitem[{{Ohlmann} {et~al.}(2016{\natexlab{b}}){Ohlmann}, {R\"opke}, {Pakmor},
  {Springel}, \& {M\"uller}}]{ohlmann2016b}
{Ohlmann}, S.~T., {R\"opke}, F.~K., {Pakmor}, R., {Springel}, V., \&
  {M\"uller}, E. 2016{\natexlab{b}}, \mnras, 462, L121

\bibitem[{{Paczynski}(1976)}]{paczynski1976a}
{Paczynski}, B. 1976, in IAU Symposium, Vol.~73, Structure and Evolution of
  Close Binary Systems, ed. P.~{Eggleton}, S.~{Mitton}, \& J.~{Whelan}, 75

\bibitem[{{Pakmor} {et~al.}(2011){Pakmor}, {Bauer}, \&
  {Springel}}]{pakmor2011d}
{Pakmor}, R., {Bauer}, A., \& {Springel}, V. 2011, \mnras, 418, 1392

\bibitem[{{Pakmor} \& {Springel}(2013)}]{pakmor2013b}
{Pakmor}, R. \& {Springel}, V. 2013, \mnras, 432, 176

\bibitem[{{Paxton} {et~al.}(2013){Paxton}, {Cantiello}, {Arras}, {Bildsten},
  {Brown}, {Dotter}, {Mankovich}, {Montgomery}, {Stello}, {Timmes}, \&
  {Townsend}}]{paxton2013a}
{Paxton}, B., {Cantiello}, M., {Arras}, P., {et~al.} 2013, \apjs, 208, 4

\bibitem[{{Paxton} {et~al.}(2015){Paxton}, {Marchant}, {Schwab}, {Bauer},
  {Bildsten}, {Cantiello}, {Dessart}, {Farmer}, {Hu}, {Langer}, {Townsend},
  {Townsley}, \& {Timmes}}]{paxton2015a}
{Paxton}, B., {Marchant}, P., {Schwab}, J., {et~al.} 2015, \apjs, 220, 15

\bibitem[{{Perets} {et~al.}(2010){Perets}, {Gal-Yam}, {Mazzali}, {Arnett},
  {Kagan}, {Filippenko}, {Li}, {Arcavi}, {Cenko}, {Fox}, {Leonard}, {Moon},
  {Sand}, {Soderberg}, {Anderson}, {James}, {Foley}, {Ganeshalingam}, {Ofek},
  {Bildsten}, {Nelemans}, {Shen}, {Weinberg}, {Metzger}, {Piro}, {Quataert},
  {Kiewe}, \& {Poznanski}}]{perets2010a}
{Perets}, H.~B., {Gal-Yam}, A., {Mazzali}, P.~A., {et~al.} 2010, \nat, 465, 322

\bibitem[{{Podsiadlowski}(1993)}]{podsiadlowski1993a}
{Podsiadlowski}, P. 1993, in Astronomical Society of the Pacific Conference
  Series, Vol.~36, Planets Around Pulsars, ed. J.~A. {Phillips}, S.~E.
  {Thorsett}, \& S.~R. {Kulkarni}, 149--165

\bibitem[{{Podsiadlowski}(2001)}]{podsiadlowski2001a}
{Podsiadlowski}, P. 2001, in Astronomical Society of the Pacific Conference
  Series, Vol. 229, Evolution of Binary and Multiple Star Systems, ed.
  P.~{Podsiadlowski}, S.~{Rappaport}, A.~R. {King}, F.~{D'Antona}, \&
  L.~{Burderi}, 239

\bibitem[{{Prust} \& {Chang}(2019)}]{prust2019a}
{Prust}, L.~J. \& {Chang}, P. 2019, \mnras, 486, 5809

\bibitem[{{Reichardt} {et~al.}(2019){Reichardt}, {De Marco}, {Iaconi}, {Tout},
  \& {Price}}]{reichardt2019a}
{Reichardt}, T.~A., {De Marco}, O., {Iaconi}, R., {Tout}, C.~A., \& {Price},
  D.~J. 2019, \mnras, 484, 631

\bibitem[{{Rogers} \& {Nayfonov}(2002)}]{rogers2002a}
{Rogers}, F.~J. \& {Nayfonov}, A. 2002, \apj, 576, 1064

\bibitem[{{Rogers} {et~al.}(1996){Rogers}, {Swenson}, \&
  {Iglesias}}]{rogers1996a}
{Rogers}, F.~J., {Swenson}, F.~J., \& {Iglesias}, C.~A. 1996, \apj, 456, 902

\bibitem[{{Schaffenroth} {et~al.}(2015){Schaffenroth}, {Barlow}, {Drechsel}, \&
  {Dunlap}}]{schaffenroth2015a}
{Schaffenroth}, V., {Barlow}, B.~N., {Drechsel}, H., \& {Dunlap}, B.~H. 2015,
  \aap, 576, A123

\bibitem[{{Schaffenroth} {et~al.}(2019){Schaffenroth}, {Barlow}, {Geier},
  {Vu{\v{c}}kovi{\'c}}, {Kilkenny}, {Wolz}, {Kupfer}, {Heber}, {Drechsel},
  {Kimeswenger}, {Marsh}, {Wolf}, {Pelisoli}, {Freudenthal}, {Dreizler},
  {Kreuzer}, \& {Ziegerer}}]{schaffenroth2019a}
{Schaffenroth}, V., {Barlow}, B.~N., {Geier}, S., {et~al.} 2019, \aap, 630, A80

\bibitem[{{Schaffenroth} {et~al.}(2014){Schaffenroth}, {Classen}, {Nagel},
  {Geier}, {Koen}, {Heber}, \& {Edelmann}}]{schaffenroth2014a}
{Schaffenroth}, V., {Classen}, L., {Nagel}, K., {et~al.} 2014, \aap, 570, A70

\bibitem[{{Schaffenroth} {et~al.}(2018){Schaffenroth}, {Geier}, {Heber},
  {Gerber}, {Schneider}, {Ziegerer}, \& {Cordes}}]{schaffenroth2018a}
{Schaffenroth}, V., {Geier}, S., {Heber}, U., {et~al.} 2018, \aap, 614, A77

\bibitem[{{Schneider} {et~al.}(2019){Schneider}, {Ohlmann}, {Podsiadlowski},
  {R{\"o}pke}, {Balbus}, {Pakmor}, \& {Springel}}]{schneider2019a}
{Schneider}, F. R.~N., {Ohlmann}, S.~T., {Podsiadlowski}, P., {et~al.} 2019,
  \nat, 574, 211

\bibitem[{{Siess} \& {Livio}(1999)}]{siess1999a}
{Siess}, L. \& {Livio}, M. 1999, \mnras, 308, 1133

\bibitem[{{Soker}(1998)}]{soker1998a}
{Soker}, N. 1998, \aj, 116, 1308

\bibitem[{{Springel}(2010)}]{springel2010a}
{Springel}, V. 2010, \mnras, 401, 791

\bibitem[{{Stephan} {et~al.}(2020){Stephan}, {Naoz}, {Gaudi}, \&
  {Salas}}]{stephan2020a}
{Stephan}, A.~P., {Naoz}, S., {Gaudi}, B.~S., \& {Salas}, J.~M. 2020, \apj,
  889, 45

\bibitem[{{V{\"o}lschow} {et~al.}(2014){V{\"o}lschow}, {Banerjee}, \&
  {Hessman}}]{volschow2014a}
{V{\"o}lschow}, M., {Banerjee}, R., \& {Hessman}, F.~V. 2014, \aap, 562, A19

\bibitem[{{Vu{\v{c}}kovi{\'c}} {et~al.}(2016){Vu{\v{c}}kovi{\'c}},
  {{\O}stensen}, {N{\'e}meth}, {Bloemen}, \& {P{\'a}pics}}]{vuckovic2016a}
{Vu{\v{c}}kovi{\'c}}, M., {{\O}stensen}, R.~H., {N{\'e}meth}, P., {Bloemen},
  S., \& {P{\'a}pics}, P.~I. 2016, \aap, 586, A146

\bibitem[{{Wilson} \& {Nordhaus}(2020)}]{wilson2020a}
{Wilson}, E.~C. \& {Nordhaus}, J. 2020, arXiv e-prints, arXiv:2006.09360

\end{thebibliography}
\end{document}